\documentclass[]{aa}
\usepackage{natbib,graphicx,fontenc}
\bibpunct{(}{)}{;}{a}{}{,}

\DeclareUnicodeCharacter{00A0}{ }

\def\apjl{ApJ}

\def\apjs{ApJS}

\def\aap{A\&A}

\def\src{XSS J12270--4859}

\def\xmm{{\it XMM-Newton}}
\def\1023{PSR J1023+0038}
\def\src{IGR J17511--3057}
\def\swift{\it Swift}
\def\inte{\it INTEGRAL}
\def\igr{\it INTEGRAL}
\def\xrt{{\it Swift}-XRT}
\def\fluxcgs{~erg~cm$^{-2}$~s$^{-1}$}
\def\lumcgs{~erg~s$^{-1}$}

\begin{document}

\title{The 2015 outburst of the accreting millisecond pulsar {\src} as
  seen by {\igr}, {\swift} and {\xmm}} \subtitle{}

\author{A.~Papitto\inst{\ref{inst0},\ref{inst1}}\thanks{alessandro.papitto@oa-roma.inaf.it} \and
  E.~Bozzo\inst{\ref{inst2}} \and C.~Sanchez-Fernandez\inst{\ref{inst3}} \and P.~Romano\inst{\ref{inst4}} \and D.~F.~Torres\inst{\ref{inst1}}$^{,}$\inst{\ref{inst5}} \and C.~Ferrigno\inst{\ref{inst2}} \and J.~J.~E.~Kajava\inst{\ref{inst3}} \and E.~Kuulkers\inst{\ref{inst3}}  }

\institute{INAF, Osservatorio Astronomico di Roma, via di Frascati 33, I-00044, Monte Porzio Catone (Roma), Italy\label{inst0} \and
  Institut de Ci\`encies de l'Espai (IEEC-CSIC), Campus UAB, Carrer de Can Magrans s/n, E-08193 Barcelona, Spain\label{inst1} \and ISDC Data Centre for Astrophysics, Chemin
  d'Ecogia 16, CH-1290 Versoix, Switzerland \label{inst2} \and
  European Space Astronomy Centre (ESA/ESAC), Science Operations
  Department, 28691 Villanueva de la Cañada, Madrid, Spain \label{inst3}\and INAF,
  Istituto di Astrofisica Spaziale e Fisica Cosmica - Palermo, Via
  U.\ La Malfa 153, I-90146 Palermo, Italy\label{inst4} \and
  Instituci\'o Catalana de Recerca i Estudis Avan\c{c}ats (ICREA),
  08010 Barcelona, Spain\label{inst5}}

\date{Received date }

\abstract{We report on {\igr}, {\swift} and {\xmm} observations of
  {\src} performed during the outburst that occurred between March 23
  and April 25, 2015. The source reached a peak flux of
  $0.7(2)\times10^{-9}$~{\fluxcgs} and decayed to quiescence in
  approximately a month. The X-ray spectrum was dominated by a
  power-law with photon index between 1.6 and 1.8, which we
  interpreted as thermal Comptonization in an electron cloud with
  temperature $> 20$ keV . A broad ($\sigma\simeq1$~keV) emission line
  was detected at an energy ($E=6.9^{+0.2}_{-0.3}$ keV) compatible
  with the K-$\alpha$ transition of ionized Fe, suggesting an origin
  in the inner regions of the accretion disk. The outburst flux and
  spectral properties shown during this outburst were remarkably
  similar to those observed during the previous accretion event
  detected from the source in 2009. Coherent pulsations at the pulsar
  spin period were detected in the {\xmm} and {\igr} data, at a
  frequency compatible with the value observed in 2009. Assuming that
  the source spun up during the 2015 outburst at the same rate
  observed during the previous outburst, we derive a conservative
  upper limit on the spin down rate during quiescence of
  $3.5\times10^{-15}$~Hz~s$^{-1}$. Interpreting this value in terms of
  electromagnetic spin down yields an upper limit of
  $3.6\times10^{26}$~G~cm$^{3}$ to the pulsar magnetic dipole
  (assuming a magnetic inclination angle of $30^{\circ}$).  We also
  report on the detection of five type-I X-ray bursts (three in the
  {\xmm} data, two in the {\igr} data), none of which indicated
  photospheric radius expansion.}

\keywords{pulsars: general --- stars: neutron -–- X-rays: binaries --- X-rays: individual: {\src}}

\titlerunning{The 2015 outburst of {\src}}\authorrunning{A.~Papitto et al.}

\maketitle

\section{Introduction}

Accreting millisecond pulsars (AMSPs hereafter) are neutron stars (NS)
that accrete matter transferred from a low mass ($M_2\la M_{\odot}$)
companion star \citep{wijnands1998}; their magnetospheres are able to
truncate the disk in-flow and channel the in-falling matter to the
regions of the surface close to the magnetic poles, producing coherent
pulsations in the X-ray light curve. The extremely quick rotation of
AMSPs is attained during a previous Gyr-long phase of sustained mass
accretion, and these sources are considered the most immediate
progenitors of millisecond radio pulsars
\citep{bisnovatyikogan1974,alpar1982,radhakrishnan1982,archibald2009,papitto2013nat}. So
far, accretion driven coherent pulsations have been detected from 17
transient low-mass X-ray binaries \citep[see, e.g.][for a
  review]{patruno2012c}. Pulsations from 15 of these objects have been
observed during relatively bright ($L_X\approx \mbox{few}\times
10^{36}$ erg s$^{-1}$) X-ray outbursts lasting few-weeks to
few-months. In two cases, PSR J1023+0038 and XSS J12270-4859
\citep{archibald2015,pap2015}, the coherent signal was detected when
the source was at a much lower luminosity level ($L_X\approx
\mbox{few}\times 10^{33}$ erg s$^{-1}$); these two sources were also
detected as radio pulsars during X-ray quiescence
\citep{archibald2009,roy2015}, similar to the AMSP IGR J18245--2452
\citep{papitto2013nat}.

{\src} was discovered by {\igr} during an outburst in September 2009
\citep{baldovin2009,bozzo2010}. The detection of 4.1 ms coherent
pulsations in the {\it Rossi X-ray Timing Explorer} light curve
allowed the identification of the source as an AMSP in a binary system
with a 3.47 hr orbital period \citep{markwardt2009}. The measured
pulsar mass function indicated a main-sequence companion star with a
mass between 0.15 and 0.44 $M_{\odot}$ \citep{papitto2010}. At
discovery, 18 type-I X-ray bursts were observed from the pulsar
\citep{altamirano2010,bozzo2010,papitto2010,falanga2011}. Evidence of
photospheric radius expansion was not displayed by any of these
bursts, and an upper limit to the source distance (6.9 kpc) was
provided by \citet{altamirano2010}.

A new outburst of {\src} was detected by {\igr} on March 23, 2015
\citep{bozzo2015,bozzo2015b}. Here, we report on a series of {\igr} and
      {\swift} observations performed throughout the event, as well as
      on a {\xmm} Target of Opportunity observation performed three
      days after the onset of the outburst.

\section{Observations}

\subsection{\swift}

The {\xrt} \citep[][]{Burrows2005:XRT} observed \src\ from March 23,
2015 at 14:08:39 (UTC) for a total of 19 ks.  A log of all available
observations for this outburst is reported in Table~\ref{tab:log}.
{\xrt} data, that were collected in both windowed-timing (WT) and
photon-counting (PC) mode depending on the source count rate, were
processed and analysed using the standard software ({\sc FTOOLS}
v6.17), calibration (CALDB 20150721), and methods.  The WT data were
never affected by pile-up; on the contrary, the PC data were
corrected, when required, by adopting standard procedures
\citep[][]{vaughan2006:050315}. Therefore WT source events were
generally extracted from circular regions of 20 pixels (1 pixel
$\simeq 2.36$\arcsec), while PC source events from annuli with outer
radius of 20 pixels and inner radius ranging from 3 to 7 pixels,
depending on the severity of pile-up. Background events were extracted
from nearby source-free regions.  A spectrum was accumulated for each
observation and each observing mode, throughout the campaign, with the
exception of the last four observations (036--039), for which a single
spectrum was extracted to increase statistics.  The data were binned
to ensure at least 20 counts per energy bin and were fit in the
0.3--10\,keV energy range. The latest responses within CALDB were
used.  Spectral modelling was performed using \textsc{xspec} v.12.8.2.

\begin{table*} 	
 \begin{center} 	
 \caption{Log of the {\swift} and {\xmm} observations considered in this paper\label{tab:log}} 	
 \label{} 	
 \begin{tabular}{ 	
l	l	l	l	l
  } 
 \hline 
 \hline 
 \noalign{\smallskip} 
 Sequence   & Obs/Mode  & Start time  (UT)  & End time   (UT) & Exposure  \\ 
           &           & (yyyy-mm-dd hh:mm:ss)  & (yyyy-mm-dd hh:mm:ss)  &(s)    \\
 & 	 & 	 & 	 & 	    \\
 & 	 & 	 & 	 & 	    \\
  Swift & & & & \\
 \hline 
 \noalign{\smallskip} 
00031492020	&	XRT/WT	&	2015-03-23 14:08:39	&	2015-03-23 16:18:57	&	860	 \\
00031492021	&	XRT/WT	&	2015-03-24 16:00:13	&	2015-03-24 16:15:58	&	926	 \\
00031492022	&	XRT/PC	&	2015-03-25 01:48:43	&	2015-03-25 06:39:55	&	923	 \\

00031492024	&	XRT/WT	&	2015-03-28 15:54:29	&	2015-03-28 16:05:58	&	682	 \\
00031492025	&	XRT/WT	&	2015-03-28 19:16:09	&	2015-03-28 19:20:58	&	266	 \\
00031492026	&	XRT/WT	&	2015-03-29 20:30:24	&	2015-03-29 20:46:58	&	981	 \\
00031492027	&	XRT/WT	&	2015-04-02 04:30:10	&	2015-04-02 06:11:26	&	97	 \\
00031492027	&	XRT/PC	&	2015-04-02 04:31:20	&	2015-04-02 06:20:54	&	1382	 \\
00031492028	&	XRT/PC	&	2015-04-04 13:33:42	&	2015-04-04 13:38:28	&	266	 \\
00031492029	&	XRT/WT	&	2015-04-06 13:52:14	&	2015-04-06 15:11:08	&	186	 \\
00031492029	&	XRT/PC	&	2015-04-06 13:52:49	&	2015-04-06 15:16:55	&	1241	 \\
00031492031	&	XRT/PC	&	2015-04-07 13:41:39	&	2015-04-07 13:57:55	&	973	 \\
00031492032	&	XRT/PC	&	2015-04-08 02:35:18	&	2015-04-08 02:50:56	&	920	 \\
00031492033	&	XRT/PC	&	2015-04-08 10:11:57	&	2015-04-08 11:58:54	&	1449	 \\
00031492034	&	XRT/PC	&	2015-04-12 00:58:10	&	2015-04-12 05:45:56	&	1316	 \\
00031492035	&	XRT/PC	&	2015-04-14 09:55:08	&	2015-04-14 10:19:55	&	1487	 \\
00031492036	&	XRT/PC	&	2015-04-16 10:00:40	&	2015-04-16 10:24:55	&	1449	 \\
00031492037	&	XRT/PC	&	2015-04-18 21:23:41	&	2015-04-18 23:04:55	&	757	 \\
00031492038	&	XRT/PC	&	2015-04-22 09:53:57	&	2015-04-22 10:06:54	&	777	 \\
00031492039	&	XRT/PC	&	2015-04-25 04:47:13	&	2015-04-25 17:56:54	&	2038	 \\
  \noalign{\smallskip}
  \hline
\noalign{\smallskip} \noalign{\smallskip}
{\xmm} & & & & \\
 \hline 
 \noalign{\smallskip} 
0770580301 & EPIC-pn & 2015-03-26 22:51:47 & 2015-03-27 19:11:33 & 73186 \\
           & RGS1    & 2015-03-26 22:16:52 & 2015-03-27 19:07:38 & 75045 \\
           & RGS2    & 2015-03-26 22:17:00 & 2015-03-27 19:08:29 & 75089 \\
 \noalign{\smallskip}
\hline

  \end{tabular}
  \end{center}
\end{table*}

\subsection{\xmm}
\label{sec:xmm}
{\xmm} \citep{jansen2001} observed {\src} for 76 ks starting on March
26, 2015 at 22:16:52 (UTC). A log of the observations analyzed is
given in Table~\ref{tab:log}. The data were reduced using SAS
v.14.0.0.

The EPIC-pn camera was operated in timing mode to achieve a time
resolution of $29.5\,\mu$s, and was equipped with a medium optical
blocking filter.  We removed soft proton flaring episodes
characterized by an EPIC-pn 10-12 keV count rate exceeding 0.8 c/s;
this reduced the effective exposure to 55.6 ks. In timing mode, the
spatial information along one of the optical axis is lost to allow a
faster readout. The maximum number of counts was recorded in pixels
characterized by RAWX coordinate 37 and 38. To extract the source
emission we considered a 21 pixel wide region (1 pixel $\simeq
4.1$\arcsec), spanning from RAWX=27 to 47. Background was extracted
far from the source, in a 3 pixel-wide region centered on
RAWX=4. Spectra were accumulated using single and double event
patterns, and re-binned with the tool \textsc{specgroup} to have at
least 25 counts per channel, and not more than three bins per
resolution element.  For the spectral extraction we followed the
recommendations of the latest calibration document on the spectral
accuracy of EPIC-pn in fast modes (Smith et
al. 2015\footnote{\url{http://xmm2.esac.esa.int/docs/documents/CAL-TN-0083.pdf}})
and applied the special gain, together with the rate dependent CTI
corrections.

Three type-I X-ray bursts were detected during the {\xmm}
observation. To analyse the {\it persistent} emission we created good
time intervals (GTIs) that eliminated a time interval starting 10~s
before and ending 150 s after each of these bursts.  The burst
  onset was identified as the first 1 s-long bin of the light curve
  that exceeded by more than 5$\sigma$ the average {\it persistent}
  countrate of 65.6 c/s.  Apart from the bursts, no evident
variability trend was seen over time scales going from a few seconds
to the length of the observation. At the average {\it persistent}
count rate observed photon pile up is not expected to affect
significantly the spectral response of the EPIC-pn in timing mode
(Smith et al.~2015).

The Reflection Grating Spectrometers (RGS) was operated in standard
spectroscopy mode. We extracted spectra from the first order of
diffraction, in which a count rate of 5.0 and 5.9 c/s was observed by
the RGS1 and RGS2 cameras, respectively. We removed the same time
intervals characterized by high flaring background and type-I X-ray
bursts as it was done for the EPIC-pn.

The EPIC MOS1 and MOS2 cameras operated in Large Window and Timing
mode, respectively. We did not consider these data in the analysis
presented here as $\simeq21\%$ and $\simeq12\%$ of the photons
detected by the two cameras respectively suffered from pile-up
distortion of the spectral response, as evaluated with the
\textsc{epatplot} tool. 

\subsection{\inte }
\label{sec:integral}

\src\ was observed within the field of view (FoV) of the instruments
on-board {\inte} from satellite revolution 1517 (starting on 2015 March
3) to 1533 (ending on 2015 April 25).

We analyzed all the publicly available {\inte} data and the data for
which we obtained data rights on the source during {\inte} AO12 using
version 10.1 of the Off-line Scientific Analysis software (OSA)
distributed by the {\inte} Science Data Center
\citep[ISDC;][]{cour03}. The {\inte} data on the source were
accumulated during observations of the Galactic Center, which are
executed following a rectangular pattern (5x5 pointings, or science
windows, SCWs) around the Galactic Center position, with typical
duratins of 2-3\,ks and a 2.17 degree step between successive
pointings. Only SCWs in which the source was located within
4.5$^{\circ}$ from the center of the JEM-X FoV \citep{lund03} were
included in the analysis of the data from this instrument. For
IBIS/ISGRI, we included all SCWs where the source was located within
12$^{\circ}$ from aim point of the instrument in order to avoid
calibration uncertainties (as suggested in the OSA manual).  We first
extracted both the IBIS/ISGRI \citep{ubertini03,lebrun03} mosaics in
the 20-80~keV and 80-100~keV energy band and noticed that there was no
significant detection of the source in the hard band. {\src} was
significantly detected in both the JEM-X mosaics extracted in the
3-10~keV and 10-20~keV energy bands.  The ISGRI and JEM-X lightcurves
of the source were extracted with a time resolution of 1 SCW in the
20-80~keV and 3-20~keV, respectively and later rebinned, to improve
the statistics, with a time resolution of 2 hours (see panels (c) and
(d) in Fig.~\ref{fig:lc}).

We accumulated a single spectrum of the source for each of the two
JEM-X units and for IBIS/ISGRI using the data in revolution 1522,
between 2015 March 25 at 03:55:44, and March 26 at 23:14:27, leading
to an effective on-source exposure time of 31.2 ks and 24.3 ks,
respectively. Such a time interval was chosen because it is the
closest to the {\xmm} observation, having an overlap of 4.5 ks with
it. In particular, we selected for the extraction of this spectrum
only the SCWs were the source was detected at more than 5$\sigma$ and
within an off-axis angle of 4.5$^{\circ}$ in order to optimize the
signal-to-noise ratio and have a fully simultaneous JEM-X+ISGRI
spectrum. The spectra response matrices were generated using the
standard energy binning of 16 channels for JEM-X and 13 channels for
IBIS/ISGRI.  The fit to these spectra was carried out simultaneously
with the \xmm\ spectrum and discussed in Sec.~\ref{sec:xmm}.

We also extracted the JEM-X1 and JEM-X2 lightcurves with a time
resolution of 2~s to search for type-I X-ray bursts, and identified
two, none in the time window overlapping with {\xmm}
  exposure. The properties of these bursts are discussed in
Sec.~\ref{sec:burst}. On the other hand, we could not identify any
burst in the ISGRI light curve.

\section{Results}
\subsection{The persistent emission}
\label{sec:persistent}

\begin{figure}
 \resizebox{\hsize}{!} 
{\includegraphics{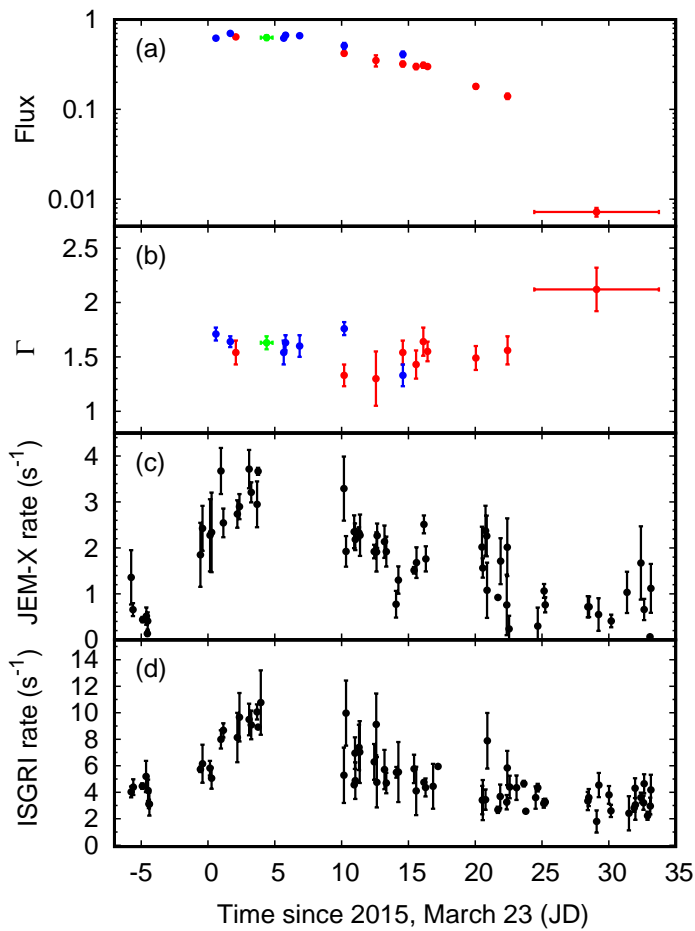}}
\caption{0.5--10 keV unabsorbed flux (in units of $10^{-9}$ erg
  cm$^{-2}$ s$^{-1}$; panel a) and photon index of the best fitting
  power-law (panel b) measured by the {\xrt} in WT (blue
  points) and PC mode (red points), and by {\xmm}
  (green point). JEM-X (panel c) and ISGRI (panel d) light curves
  were binned with a time resolution of 2 hours. }
 \label{fig:lc}
\end{figure}

The {\xrt} 0.3--10 keV spectra were modelled simultaneously with an
absorbed power-law, forcing the absorption column to take the same
value in every observation. This resulted in an estimate of the
absorption column density (described with the model \textsc{phabs} in
\textsc{xspec}) of $N_H=1.18(4)\times10^{22}$ cm$^{-2}$. The chi
squared of the fit was 1632 for 1411 degrees of freedom.  The
probability of obtaining a fit chi-squared as high or higher if the
data are drawn from the model distribution is $3.5\times10^{-5}$,
smaller than the significance level of $2.7\times 10^{-3}$ we set to
determine the goodness of the fit. s probability always larger than
the significance level adopted. Letting the absorption column as
  a free parameter in different observations did not yield a
  statistically significant decrease of the fit chi-squared. However,
we did not find any evident unmodelled residuals that motivated the
addition of more spectral components, and the chi squared of the
individual spectra resulted in a null hypothesis probability always
larger than the significance level adopted.  Systematic errors we did
not take into account possibly yielded the relatively large residuals
of the simultaneous modelling of the {\xrt} spectra with an absorbed
power law.  The 0.5--10 keV unabsorbed flux attained a maximum of
$0.7(2)\times10^{-9}$ erg cm$^{-2}$ s$^{-1}$ on March 24, during
observation 00031492021 (see Table~\ref{tab:log}). The flux decay
started around March 29 and was characterized by a roughly linear
decay, until the flux decreased abruptly on April 16 (i.e. 25 days
after the first detection of the outburst) to a level $\simeq100$
fainter than the outburst peak
($F(0.5-10\,\mbox{keV})=7.2(8)\times10^{-12}$ erg~cm$^{-2}$~s$^{-1}$
measured coadding the observations performed between April 16 and
April 25). The photon index of the power-law that fitted the {\xrt}
spectra laid between $1.3$ and $1.6$, becoming much steeper
($\Gamma=2.1(2)$) in the last group of observations performed when the
source had already fainted significantly.  The values of the
unabsorbed 0.5--10 keV flux and power law index are plotted in the
upper panels (a and b) of Fig.~\ref{fig:lc} using blue and red points
for observations performed in WT and PC modes, respectively.

\begin{figure}
 \resizebox{\hsize}{!}
           {\includegraphics{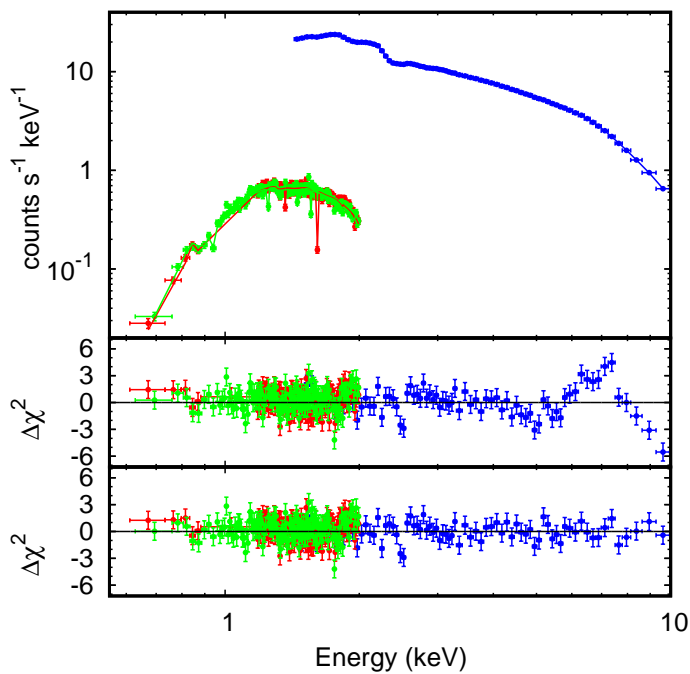}}
\caption{ Top panel shows the X-ray spectrum of {\src} observed by
    the EPIC-pn (blue points), RGS1 (red points), and RGS2 (green
    points) during the {\xmm} observation.  Spectra were rebinned for
    graphical purposes. The bottom and middle panels show residuals
    with respect to the model listed in the central column of
    Table~\ref{tab:spectrum} (see text for details), with and without
    the addition of an emission line centered at
    $E_1=6.9^{+0.2}_{-0.3}$~keV to the model, respectively.}
 \label{fig:xmmspectrum}
\end{figure}

The spectra observed on March 26, 2015 by the EPIC-pn (1.4--11
keV)\footnote{EPIC-pn data below 1.4 keV were discarded to avoid the
  spurious soft excess that sometimes appears in the EPIC-pn spectra
  obtained in fast modes, and of which we found hints also in this
  observation \citep[see, e.g.][]{hiemstra2011}.}  and the two RGS
cameras (0.6--2.0 keV) on-board {\xmm} were modelled simultaeously by
treating the normalization of the two RGS spectra with respect to the
EPIC-pn as free parameters (see top panel of
  Fig.~\ref{fig:xmmspectrum}). We used a thermal Comptonization model
\citep[\textsc{nthcomp},][]{zdziarski1996,zycki1999} modified by the
interstellar absorption. No high-energy cut-off was detected in the
{\xmm} energy band and we fixed the temperature of the Comptonizing
electron cloud to $kT_e=30$~keV, in line with the results obtained
modelling the INTEGRAL spectrum observed immediately before the XMM
pointing (see below).  The Comptonization component was described by
an asymptotic photon index $\Gamma=1.62^{+0.03}_{-0.11}$, and a
temperature of soft input photons of $kT_{s}<0.98$~keV, yielding a
high value of the fit chi-squared of 1963.3 for 1286 degrees of
freedom. To account for the residuals at soft ($kT\la$1.5 keV)
energies we added a single-temperature black-body component
(\textsc{bbodyrad} in \textsc{xspec}) with
$kT_\textrm{bb}=0.59^{+0.15}_{-0.11}$~keV, and a a disk emission
component (\textsc{diskbb} in \textsc{xspec}) with temperature of
$kT_{in}=0.26^{+0.04}_{-0.03}$~keV. The addition of these two
components yielded a significant decrease of the $\chi^2$ by
$\Delta\chi^2=-74.3$ (giving $\chi^2=1888.9$ for 1284 d.o.f.) and
-124.6 (giving $\chi^2=1764.29$ for 1282 d.o.f.), for the addition of
two degrees of freedom each, respectively. We evaluated with an F-test
that such improvements of the model $\chi^2$ are equivalent to a
probability of $<10^{-11}$ that they are due to chance.  The
normalizations of these two thermal components indicate apparent radii
of $R_{in}/(\cos{i})^{1/2}=(34\pm20)\,d_{6.9}$ km and
$R_{bb}=3.2^{+4.6}_{-1.0}\,d_{6.9}$ km for the disk and the blackbody
component, respectively. Here, $d_{6.9}$ is the distance to the source
in units of 6.9 kpc. Local residuals of the EPIC-pn data at energies
of $1.8$, $2.2$ and $6.9$ keV were modelled using three Gaussian
emission features. The first two features are narrow, and their
energies are compatible with calibration residuals of the Si K and Au
M edges, known to frequently affect the EPIC-pn spectra in timing
mode. The feature at higher energy had a centroid at
$E=6.9^{+0.2}_{-0.3}$~keV, compatible with the K$\alpha$ transition of
ionized Fe. With the inclusion of this feature the model $\chi^2$
decreased by $\Delta\chi^2=130.8$ for the addition of three free
parameters (see \citealt{protasov2002} and \citealt{stewart2009} for a
discussion of the use of the fit chi-squared to measure the
significance of spectral lines). The residuals obtained with and
  without the addition of the Fe emission line to the model are shown
  in the bottom and middle panel of Fig.~\ref{fig:xmmspectrum},
  respectively. The line width
($\sigma=1.1^{+0.4}_{-0.2}$~keV) and strength (equivelent width of
$\simeq150$~eV), suggested  broadening in the inner parts of an
accretion disk. An absorption edge at an energy compatible with the
OVIII transition ($0.871$~keV) was also added to model residuals of
the RGS spectra. With the addition of such emission features, the
reduced $\chi^2$ of the fit attained a value of 1441.8 (for 1274
degrees of freedom). The probability that a value of the $\chi^2$ as
high or higher is produced if the data are drawn from the model
distribution is $6.8\times10^{-4}$. This probability is smaller than
significance level we set ($p=2.7\times10^{-3}$), but the absence of
evident residuals, and the 2\% uncertainty quoted by Smith et
al. (2015)\footnote{http://xmm2.esac.esa.int/docs/documents/CAL-TN-0018.pdf}
for the relative effective area calibration of the EPIC-pn, motivated
us to accept this model without the addition of further components.
Model parameters of the best-fit are listed in the central column of
Tab.~\ref{tab:spectrum}.

 The broadness of the Fe line motivated us to attempt modelling it
 using a \textsc{diskline} component \citep{fabian1989}. However, the
 best fit was obtained only for extreme values of the parameters that
 control the line width ($R_{in}<8.5$ R$_g$, $i>67^{\circ}$), and did
 not yield a significant improvement of the description of the
 spectrum, as the chi-squared of the fit decreased by
 $\Delta\chi^2=3.7$ for the addition of three free parameters.

\begin{figure}
 \resizebox{\hsize}{!}
           {\includegraphics{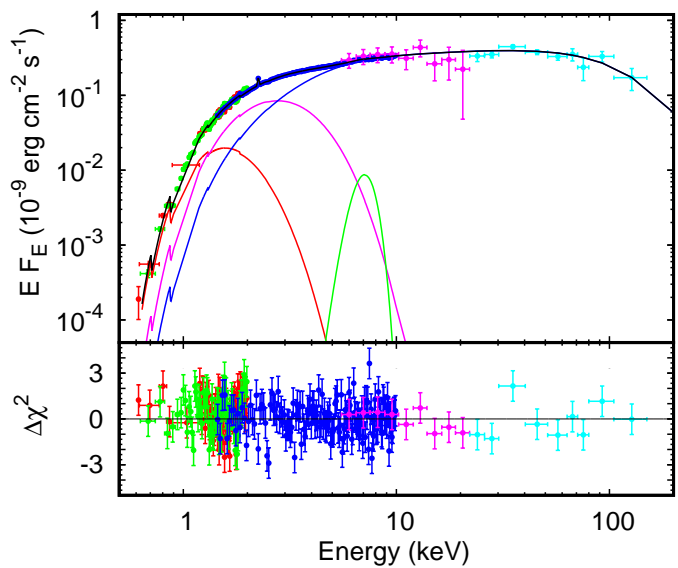}}
\caption{Top panel shows the X-ray spectrum of {\src} observed by the
  EPIC-pn (blue points), RGS1 (red points), RGS2 (green points), JEM-X
  (magenta points) and ISGRI (cyan points) during the 2015
  outburst.  RGS1 and RGS2 spectra were rebinned for graphical
  purposes. The best-fit model (black dashed line), the Fe line (green
  dashed line), the disk (red dashed line), single-temperature black
  body (magenta dashed line), and the Comptonized (blue dashed line)
  are also shown. Bottom panel shows residuals with respect to the
  best fitting model. }
 \label{fig:epnspectrum}
\end{figure}

We then fitted the {\xmm} EPIC-pn and RGS spectra together with the
{\inte} JEM-X (5-25 keV) and ISGRI (20-150 keV) spectra which were
taken on March, 26 (between 03:55:44 and 23:14:27). An AMSP like
{\src} does not usually show drastic spectral variability on
timescales of a day or less \citep{ibragimov2011}, and the analysis of
Swift data presented earlier where no significant spectral changes
were detected over the outburst, supports the simultaneous modelling
of the XMM-Newton and INTEGRAL data in spite of the limited overlap
($\simeq4.5$ ks) between these observations.  The ISGRI spectrum
allowed us to weakly constrain the electron temperature of the
Comptonized component as $kT_e=33^{+69}_{-11}$ keV, while the rest of
the spectral parameters are basically unchanged with respect to those
determined by {\xmm} spectra alone. The spectrum and residuals from
the best-fit are plotted in Fig.~\ref{fig:epnspectrum}, while model
parameters are listed in the rightmost column of
Tab.~\ref{tab:spectrum}.

We also considered the addition to the model of a component describing
the reflection of the Comptonizing photon spectrum off the accretion
disk. We removed the Gaussian describing the Fe emission line, and
added a \textsc{reflionx} \citep{ross2005} component to the model,
convolved with relativistic disk blurring (\textsc{rdblur},
\citealt{fabian1989}).  The parameters of the disk reflecting surface
were held fixed to inner and outer radii of $R_{in}=10$~R$_g$ and
$R_{out}=1000$~R$_{g}$, respectively, and index of the power law
dependence of emissivity $\beta=-2$. The best fit was obtained for a
ionization parameter $\xi=1500^{+1000}_{-500}$, and a flux of
$0.10(5)\times10^{-10}$~{\fluxcgs} in the reflection component, equal
to $\simeq10\%$ of the irradiating flux and similar to the value
observed by \citet{papitto2011}. The addition of reflection did not
yield a significant decrease of the model $\chi^2$ ($\chi2=1457.1$ for
1291 degrees of freedom), though, and we could not conclude that a
reflection component was significantly detected in the considered
dataset, probably because of low counting statistics.

\begin{table}
\caption{Spectral parameters of {\src}}
\label{tab}
\centering
\renewcommand{\footnoterule}{} 
\begin{tabular}{@{}l c c }
\hline
\hline
\noalign{\smallskip} 
Parameter & {\it XMM} & {\it XMM}--{\it IGR}  \\
\hline
\noalign{\smallskip} 
\vspace{0.2cm}$N_H$ (10$^{22}$~cm$^{-2}$) & $1.03^{+0.10}_{-0.07}$ & 0.99(5)\\
$\tau_{OVIII}$ & 0.18(7) &  \\
$kT_{in}$ (keV)   & $0.26^{+0.04}_{-0.03}$ & 0.17(7) \\
\vspace{0.2cm}$R_{in}/(cos{i})^{1/2}$ ($d_{6.9}$ km) & $34\pm20$ & $20^{+5}_{-7}$ \\
$kT_\textrm{bb}$ (keV)   & $0.59^{+0.16}_{-0.12}$ & $0.62(3)$\\
\vspace{0.2cm}$R_\textrm{bb}$ ($d_{6.9}$ km) & $3.2^{+4.6}_{-1.0}$ & $6.6^{+0.7}_{-0.9}$  \\
$\Gamma$ & $1.63^{+0.03}_{-0.11}$ & $1.85^{+0.22}_{-0.15}$ \\
$kT_{s}$ (keV) & $<0.98$ &  $1.3(2)$\\
$kT_e$ (keV) & $(30.0)$ & $33^{+69}_{-11}$\\
\vspace{0.2cm}$F^{nth}_{-9}$ (erg cm$^{-2}$ s$^{-1}$) & $0.49\pm0.15$  & $1.18(6)$ \\
$E_1$ (keV) & $6.89^{+0.18}_{-0.27}$ &  $6.93^{+0.23}_{-0.21}$ \\
$\sigma_1$ (keV) & $1.07^{+0.40}_{-0.22}$ & $0.78^{+0.4}_{-0.3}$ \\
\vspace{0.2cm}$N_1$ ($10^{-4}$ cm$^{-2}$ s$^{-1}$) & $5.7^{+9.6}_{-1.9}$ & $2.2^{+2.2}_{-1.1}$  \\
$E_2$ (keV) & $1.86(2)$ & $1.86(2)$ \\
\vspace{0.2cm}$N_2$ ($10^{-4}$ cm$^{-2}$ s$^{-1}$) & $1.2(4)$ & $1.2(4)$ \\
$E_3$ (keV) & $2.24(1)$ & $2.24(7)$  \\
\vspace{0.2cm}$N_3$ ($10^{-4}$ cm$^{-2}$ s$^{-1}$) & $2.7(6)$ & $2.7(3)$\\
RGS1/EPN & 0.98(1) & 0.98(1) \\
RGS2/EPN & 0.98(1) & 0.98(1) \\
JEM-X/EPN & ... & 0.58(9) \\
ISGRI/EPN & ... & 1.13$^{+0.39}_{-0.25}$ \\
$F_{-9}$ (erg cm$^{-2}$ s$^{-1}$) & $0.63(4)$  & $1.43(6)$   \\
$\chi^2$(dof) & 1441(1274) & 1457(1291) \\
\hline
\end{tabular}
\tablefoot{Best fit parameters of the spectrum observed by {\xmm} (RGS
  in the 0.6-2.0~keV energy band and EPIC-pn in the 1.4--10~keV range)
  on March 26,~2016 (central column), and of the simultaneous spectrum
  observed by {\xmm} and INTEGRAL (JEM-X in the 5--25~keV, and ISGRI
  in the {$20$--$150$~keV} band; rightmost column). Fluxes are unabsorbed,
  given in units of $10^{-9}$~{\fluxcgs}, and evaluated in the
  0.5--10~keV for the XMM spectrum, and the 0.5--100~keV energy band
  for the XMM-IGR spectrum, respectively. Uncertainties are quoted at
  the 90\% confidence level.}
\label{tab:spectrum}
\end{table}

\subsection{Timing analysis}

To perform a timing analysis of the coherent signal shown by the
source in the {\xmm} EPIC-pn data, we first converted the times of
arrival of the X-ray photons to the Solar system barycenter using the
source position determined by \citealt{nowak2009} using Chandra
observations (see also \citealt{paizis2012}). Coherent pulsations at
244.8~Hz were easily detected at an rms amplitude of $\simeq15$ per
cent in 50 s-long intervals. Starting from the orbital solution
provided by \citet{riggio2011}, we applied standard timing techniques
\citep[see, e.g.][]{papitto2011} to study the evolution of the pulse
phase computed over 500 s-long intervals. The values of the orbital
parameters we obtained, namely the semi-major axis of the NS orbit $a
\sin{i}/c$, the orbital period $P_{orb}$, the epoch of passage at the
ascending node $T^*$ and eccentricity $e$, as well as the value of the
spin frequency $\nu$, and spin frequency derivative $\dot{\nu}$, are
given in Table~\ref{tab:spin}.  The spin frequency derivative could
not be constrained over the relatively short exposure of the {\xmm}
observation. As we show in Sec.~\ref{sec:disc}, spin up torques are
expected to increase the pulsar frequency at a rate of
$\approx$~few~$\times10^{-13}$~Hz~s$^{-1}$, two orders of magnitude
lower with respect to the upper limit we measured
($|\dot{\nu}|<1.2\times10^{-11}$~Hz~s$^{-1}$, at 3$\sigma$ confidence
level). Because of the relatively short exposure, the best fit values
of the spin frequency correlated to a high degree with the
(unconstrained) spin frequency derivative. Similarly to what was done
by \citet{papitto2010} for the 2009 outburst of {\src}, we quote in
Tab.~\ref{tab:spin} the pulsar spin frequency obtained fixing
$\dot{\nu}=0$ in the fit (i.e. the average spin frequency over the
considered interval). A compatible value of the spin frequency is
obtained if a value of the spin-up rate of the order of that expected
from the observed X-ray luminosity
($\dot{\nu}=2\times10^{-13}$~Hz~s$^{-1}$) is considered, justifying
our choice.

The error on the position of the source is an additional source
  of uncertainty on the value of the spin frequency \citep[see,
    e.g.][]{lyne1990}. To evaluate the frequency shift produced by a
  difference between the actual value of the ecliptic coordinates of
  the source and that used to report X-ray photons to the Solar system
  barycenter, we used  \citep[see, e.g., Eq.~4 in][]{papitto2011}:
\begin{equation}
\label{eq:poserr}
\delta\nu_{pos}=\nu\,y\frac{2\pi}{P_{\oplus}}[\cos{M_i}\cos{\beta}\,\delta\lambda+\sin{M_i}\sin{\beta}\delta\beta].
\end{equation}
Here, y is the Earth distance from the Solar System barycentre in lt-s,
$\lambda$ and $\beta$ are the ecliptic longitude and latitude,
respectively, $\delta\lambda$ and $\delta\beta$ are the respective
uncertainties, $M_i=[2\pi(T_i-T_{\gamma})/P_{\oplus}]-\lambda$, $T_i$
is the start time of observations considered, $T_{\gamma}$ is the
nearest epoch of passage at the vernal point, and P$_{\oplus}$ is the
Earth orbital period. The error of the position of the source ($0.6''$
at 90\% confidence level, \citealt{nowak2009,paizis2012}, which
translates into $\sigma_{\lambda}=1.5\times10^{-6}$~rad,
$\sigma_{\beta}=1.8\times10^{-6}$~rad) yielded an uncertainty of
$\sigma_\nu^{pos}=3\times10^{-8}$ Hz on the value of the measured spin
frequency. Adding in quadrature this uncertainty to the statistical
error gave an error of $\sigma_\nu=4\times10^{-8}$~Hz on the average
spin frequency measured by {\xmm} in 2015.

\begin{figure}
 \resizebox{\hsize}{!} 
{\includegraphics{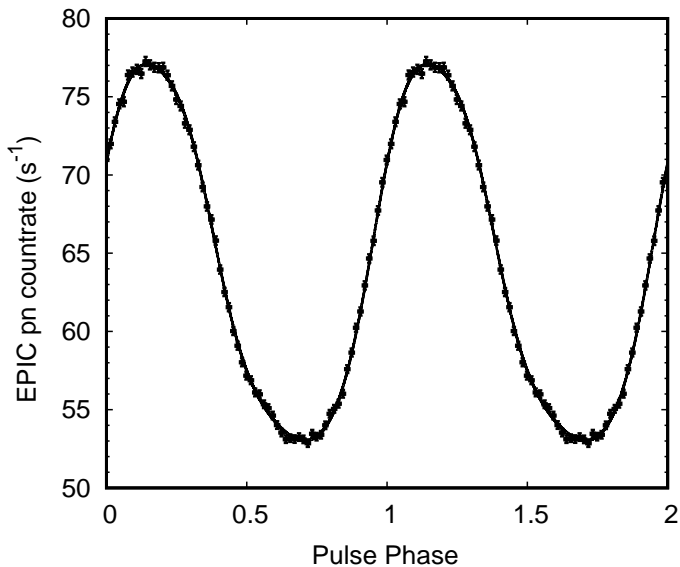}}
\caption{0.3--10 keV pulse profile of {\src} as observed by the EPIC
  pn during the 2015 outburst. The best fitting model (solid line) is
  composed by four harmonics with background-subtracted, rms
  amplitudes of 14.31(6)\%, 1.53(6)\%, 0.72(6)\% and 0.26(6)\%,
  respectively. Two cycles are displayed for clarity.  }
 \label{fig:pulse}
\end{figure}

Folding the entire EPIC-pn observation around the best fitting
value of the spin frequency, we obtained the 0.3-10 keV pulse profile
displayed in Fig.~\ref{fig:pulse}. It was successfully modelled by four
harmonic components, similar to the pulse profile observed in 2009
\citep{papitto2010}.

\begin{figure}
 \resizebox{\hsize}{!} 
{\includegraphics{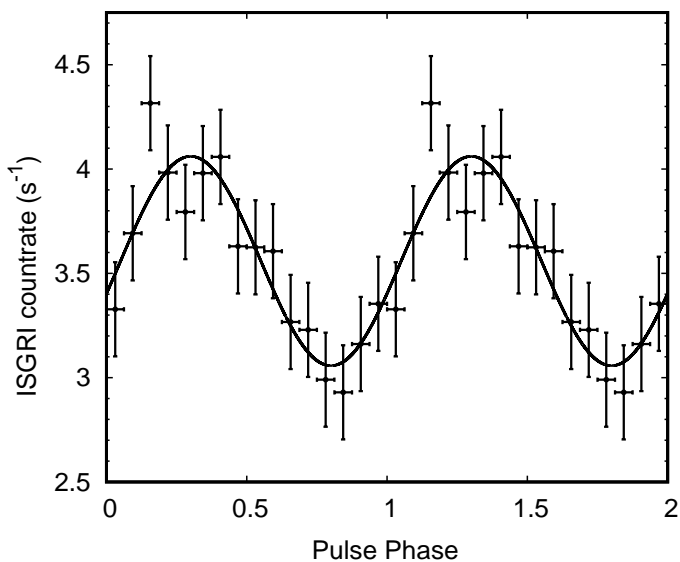}}
\caption{Background subtracted pulse profile of {\src} in the 20-100
  keV energy band extracted from 221 ks of dead-time corrected on-axis
  equivalent exposure time in the IBIS/ISGRI data (from 2015-03-22
  08:56:29 to 04-09 05:52:01 UTC); the vertical axis displays
  equivalent count per second for on-axis pointing. }
 \label{fig:igrpulse}
\end{figure}

We searched for coherent pulsations in the hard X-rays ISGRI band,
considering the data observed by ISGRI from 2015 March 22 at 08:56:29
to April 9 at 05:52:01 UTC.  We converted the arrival times of the
events recorded by ISGRI at the Solar system barycenter using the
position derived by Chandra, and then to the line of nodes using the
ephemeris listed in of Table~\ref{tab:spin}.  We employed a customary
software dedicated to phase-resolved spectroscopy of X-ray binaries
with IBIS \citep{Segreto2007} to extract the background subtracted
pulse profile in the broad 20-100 keV energy band. We detected the
signal at the pulsar spin frequency determined by {\xmm}, but the low
signal to noise ratio of the ISGRI data prevented us from obtaining an
independent measurement. The pulse profile is displayed in
Fig.~\ref{fig:igrpulse}. Fitting the profile with a constant and a
sinusoid, we determined that the ratio between the sine amplitude and
the constant as 14$\pm$3\%, in agreement with what was found by
\citet{falanga2011} during the 2009 outburst of this source. Splitting
the energy range in two bands, we obtained amplitudes of 22$\pm$6\%
and 12$\pm$3\%.for the energy ranges 20–30\,keV and 30–100\,keV,
respectively, suggesting a decrease of the pulsed emission at
higher energy.

\begin{table}
\caption{Spin and orbital parameters of {\src}}
\label{tab}
\centering
\renewcommand{\footnoterule}{} 
\begin{tabular}{@{}l c }
\hline
\hline
$<\nu>$ (Hz) & 244.83395112(3) \\
\vspace{0.2cm}$\dot{\nu}$ (Hz s$^{-1}$) & $<1.2\times10^{-11}$ \\
$a \sin{i}/c$ (lt-s) & 0.275196(4) \\
$P_{orb}$ (s) & 12487.53(2) \\
$T^*$ (MJD) & 57107.8588090(15) \\
\vspace{0.2cm}$e$ & $<8\times10^{-5}$ \\
$\chi^2$(d.o.f.) & 141.5(121) \\
\hline
\end{tabular}
\label{tab:spin}
\tablefoot{The timing solution is referred to the epoch
  $T_{0}=57107.954157$ MJD. The error on the spin frequency does not
  take into account the uncertainty driven by the indetermination on
  the position of the source ($\sigma_\nu^{pos}=3\times10^{-8}$ Hz) }
\end{table}

\begin{figure}
 \resizebox{\hsize}{!} 
{\includegraphics{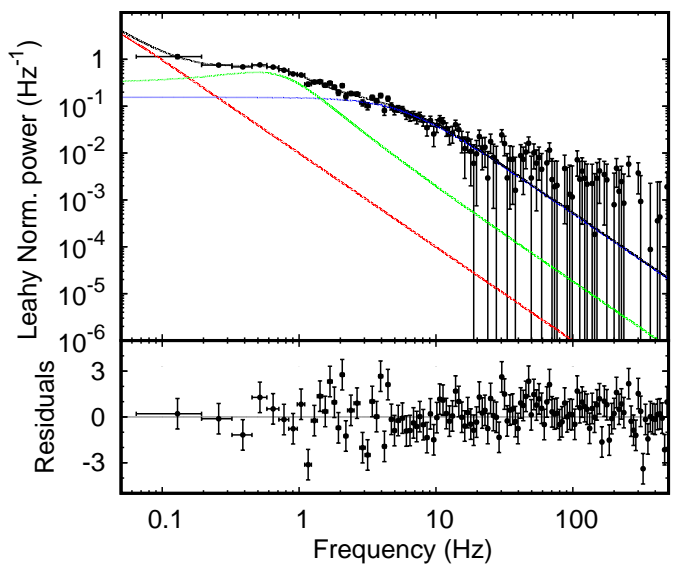}}
\caption{Average Leahy normalized power spectrum of the EPIC-pn time
  series (top panel). A counting noise constant level of $1.9993(4)$
  is subtracted to the power. The two flat-top noise components are
  plotted with a red and a blue solid line, respectively, and the
  discrete feature centered at $\nu_3=0.5\pm0.1$~Hz as a green solid
  line (see text for details). The spike at the pulsar spin frequency
  was removed for graphical purposes. Bottom panel shows residuals in
  units of the standard deviation $\sigma$.  }
 \label{fig:pds}
\end{figure}

In order to study the aperiodic time variability, we produced a power
density spectrum of the 0.5--10 keV EPIC-pn time series with 59~$\mu$s
time resolution (yielding a Nyquist frequency of 8468~Hz). We averaged
the Leahy-normalized fast Fourier transforms performed over
approximately 8 s-long intervals (for a total of 9457 spectra, each
extending down to 0.1~Hz). The resulting average power density
spectrum was rebinned as a geometrical series with a ratio of 1.02.
The 0.1--500 Hz spectrum was modelled with the sum of two flat-top
noise components modeled with Lorentizian functions centered at
$\nu=0$ with widths $W_1=(3\pm1)\times10^{-2}$~Hz and
$W_2=11.5+/-1.3$~Hz, and a discrete feature centered at
$\nu_3=0.5\pm0.1$~Hz with width $W_3=1.2\pm0.2$~Hz, giving a fit
$\chi^2$ of 611.2 for 522 degrees of freedom. We searched for kHz QPOs
fitting a Lorentzian with quality factor fixed to a value of 4, but
found no significant QPO with a 3$\sigma$ upper limit of 1$\%$ on the
rms amplitude, a value lower than that characterizing the high
frequency QPOs reported by \citet{kalamkar2011}. The average power
spectrum and residuals with respect to the best fit model are plotted
in Fig.~\ref{fig:pds}.

\subsection{Type-I X-ray bursts}
\label{sec:burst}

\begin{figure}
 \resizebox{\hsize}{!}
           {\includegraphics{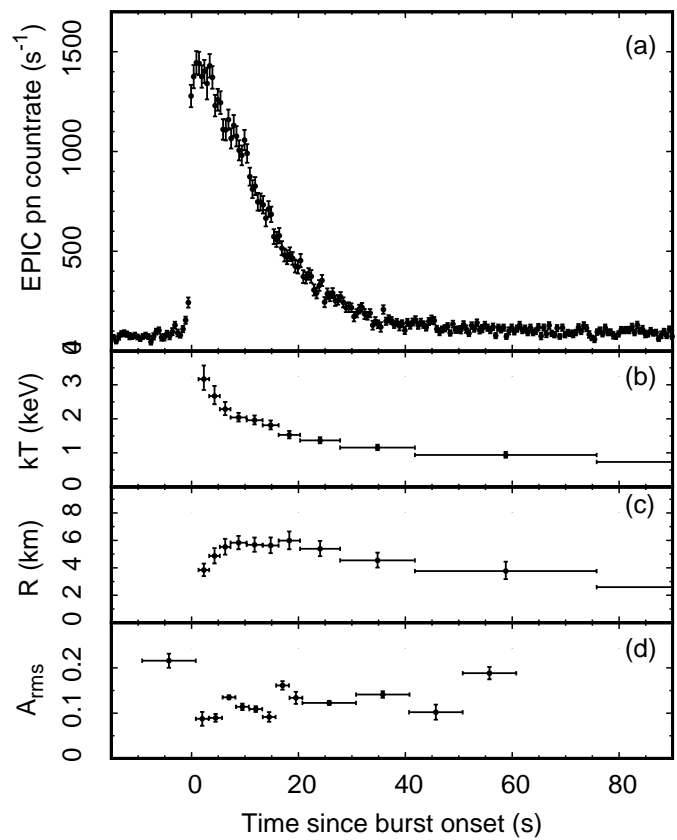}}
\caption{0.3-10 keV light curve of the first type-I X-ray burst
  oserved by {\xmm} during the 2015 observation (panel a). The burst
  onset took place at MJD 57108.01023. The temperature and the
  apparent radius (evaluated considering a distance to the source of
  6.9 kpc) are plotted in panels (b) and (c). Panel (d) shows the rms
  amplitude of pulsations at the spin frequency of the pulsar (see
  Table~\ref{tab:spin}).}
 \label{fig:xmmburst}
\end{figure}

During the {\xmm} observation, three type-I X-ray bursts were
detected, sharing very similar properties. The light curve of the
first burst is displayed in panel (a) of Fig.~\ref{fig:xmmburst}.  The
start times of the three bursts were $T_1=57108.01023(1)$~MJD,
$T_2=57108.34130(1)$~MJD, and $T_3=57108.67847(1)$~MJD, respectively.  The
burst rise times were $\approx 1$~s, and were followed by exponential
decays which started a few seconds after the maxima. The measured
decay e-folding times were $\tau_1=12.1(2)$~s, $\tau_2=12.6(2)$~s and
$\tau_3=12.4(2)$~s for the three bursts, respectively. The time
elapsed between consecutive bursts was $\Delta t_1=28.6$~ks and
$\Delta t_2=29.1$~ks.

We extracted the EPIC-pn burst spectra removing the two brightest
pixel columns to minimize the effect of pile up, and subtracted the
spectrum of the persistent emission as the background. Close to the
burst peak spectra were extracted over 2~s long intervals; at later
times the exposure was increased to ensure that the counting
statistics allowed a meaningful spectral modelling.  Spectral channels
were rebinned in the same way as the persistent spectrum.  The spectra
extracted over the different intervals were simultaneously fit by an
absorbed blackbody with variable temperature and radius, resulting in
a fit $\chi^2$ of 510.9 (for 556 degrees of freedom), 559.43 (522) and
589.16 (518) for the three bursts, respectively. The 0.5--10 keV burst
flux attained a maximum value of
$(1.5\pm0.1)\times10^{-8}$~\fluxcgs. The temperature decreased
steadily from $\simeq 3$ keV during the burst evolution (see panel b
of Fig.~\ref{fig:xmmburst}), while the apparent radius of the emitting
region peaked at $\simeq 6\,d_{6.9}$ km after ten seconds since the
burst onset, and then decayed smoothly (see panel c of
Fig.~\ref{fig:xmmburst}).

Coherent oscillations at the pulsar spin frequency were detected at a
somewhat lower rms amplitude ($\simeq10$ per cent) than that of
persistent pulsations ; see panel d of
Fig.~\ref{fig:xmmburst}). Considering that the burst emission was up
to 25 times brighter than the persistent flux, such a decrease of the
pulse amplitude could not be ascribed to the increase of the unpulsed
flux, but suggests instead that the oscillations detected during the
burst were different than the coherent signal observed during
quiescence. Pulsations were detected both close to the burst peak and
during the burst decay. The phase of burst oscillations measured over
5-s long intervals is generally consistent within the uncertainties
with the phase of the persistent profile, even if deviations up to 0.2
in phase are observed close to the burst peaks.

Two bursts were detected by {\inte}, with trigger times of
$T_4$=57116.43276(2)~MJD and 57120.11919(2)~MJD. From the JEM-X
lightcurves we measured a peak count-rates of 344.6 +/- 47.6 c/s
(3-20 keV) and 177.5 +/- 38.8 c/s (3-20 keV), respectively for the
two bursts. Because JEM-X spectra cannot be extracted for time
intervals shorter than 8~s, we estimated the peak flux by comparing
the above count-rates with that of the Crab in the same energy band as
obtained from the {\inte} calibration observations performed in the
satellite revolution 1520 (on 2015 March 19). We obtained a 3-20~keV
X-ray flux of 1.3+/-0.2~Crab and 0.7+/-0.1~Crab, respectively. These
translate into $\sim 3.0\times10^{-8}$~erg~cm$^{-2}$~s$^{-1}$ and
$\sim1.6\times10^{-8}$~erg~cm$^{-2}$~s$^{-1}$. The e-folding decay times of the
two bursts measured from the JEM-X lightcurves were  $7.3\pm0.1$~s
and $6.0\pm0.1$~s.

\section{Discussion}
\label{sec:disc}

During the outburst detected in early 2015, {\src} attained a peak
flux of $\simeq$ 0.6--0.7$\times10^{-9}${\fluxcgs} (0.5--10~keV,
unabsorbed, see Sec.~ \ref{sec:persistent}). This value is compatible
with the peak flux observed in the last outburst detected in 2009
\citep{bozzo2010}. Assuming a bolometric correction factor of $\sim2$
(estimated from the ratio between the flux estimated in the
0.5--100~keV and 0.5--10~keV energy band; see Table~2), the observed
peak flux corresponds to a 0.5-100 keV luminosity of
$\simeq8\times10^{36}\,d_{6.9}${\lumcgs}. The outburst flux decayed
during $\approx25$ d before the source became no longer detectable. The
spectral shape observed by {\xrt} was roughly constant, with a
softening of the spectrum at the end of the outburst.

Similarly to the 2009 outburst \citep{papitto2010,falanga2011,
  ibragimov2011}, and to other AMPSs \citep{patruno2012c}, the X-ray
spectrum of {\src} observed during the 2015 outburst was dominated by
a hard component described by a power-law with index
$\Gamma\simeq1.6$--$1.8$, and a cut-off at a temperature of $\ga 20$
keV. The emission observed by {\xmm} and {\igr} during the 2015
outburst was modelled as thermal Comptonization of soft photons with
temperature $kT_{s}\ga 1$ keV, off a cloud of hot electrons
($kT_e\ga20$ keV). Two thermal components were detected at soft
X-rays, and interpreted as radiation coming from an accretion disk
truncated a few tens of km from the NS surface and from the NS surface
itself.  Signatures of disk reflection such as a broadened
($\sigma\approx 1$ keV) line centered at an energy compatible with
ionized iron was also observed in the {\xmm} spectrum.

X-ray pulsations at the spin period of {\src} were detected by the
{\xmm} EPIC-pn at an RMS amplitude of $\simeq15$ per cent in the
0.5-10 keV band, and by {\igr} at an amplitude of $14$ per cent in the
20--100 keV band. Up to four harmonics were needed to model the pulse
profile observed in the soft X-ray band, similar to the case of the
2009 outburst \citep{papitto2010}. We searched for kHz quasi-periodic
oscillations as those reported by \citet{kalamkar2011}, but did not
find any hint in the Fourier power spectrum.

The relatively short exposure ($\sim$~70~ks) over which pulsations
were detected prevented us from studying the frequency evolution
driven by the accretion of matter. Only a loose upper limit on the
spin frequency derivative of $\sim10^{-11}$~Hz~s$^{-1}$ could be
obtained. On the other hand, the expected spin up rate driven by
accretion torques is:
\begin{equation}
\label{eq:nudotaccr}
\dot{\nu}\simeq\frac{1}{2\pi I_*}\frac{R_* L_X}{\sqrt{GM_*}}\sqrt{R_{in}},
\end{equation}
where $M_*$, $R_*$ and $I_*$ are the NS mass, radius, and moment of
inertia, respectively, $R_{in}$ is the disk truncation radius and
$L_X$ is the accretion luminosity. The accretion luminosity derived
from the flux observed in the 0.5-100 keV band is
$L_X=8.1\times10^{36}\,d_{6.9}$ erg s$^{-1}$. Evaluating
Eq.~\ref{eq:nudotaccr} for a disk truncated at the corotation radius
($R_c=42.8\,m_{1.4}^{1/3}$ km for {\src}), and taking $M_*=1.4$
M$_{\odot}$, $R_*=10$ km and $I=10^{45}$ g cm$^2$, yields
$\dot{\nu}\simeq2\times10^{-13}$ Hz s$^{-1}$, two orders of magnitude
lower than the upper limit we could set based on the {\xmm}
observation alone. A spin-up at an average rate of
  $\dot\nu_{09}=(1.45\pm0.16)\times10^{-13}$~Hz~s$^{-1}$ was indeed
  reported by \citet{riggio2011} for the 2009 outburst, when the
  source emitted a luminosity similar to that observed during the 2015
  outburst.

Assuming that the spin frequency observed by {\xmm} during a
$\approx70$ ks exposure is a good tracer of the actual spin frequency
of the pulsar, we can derive constraints on the evolution of the spin
frequency of the pulsar during the $\Delta t_q=1991$ d elapsed between
the end of 2009 outburst (MJD 55113; see Fig.1 of
\citealt{riggio2011}), and the beginning of the 2015 outburst (MJD
57104; \citealt{bozzo2015}). In 2009, {\xmm} measured an average spin
frequency of $\nu_{09}=244.83395121(4)$ Hz \citep{papitto2010}, and
the frequency decrease between this value and that measured in 2015
outburst is $\Delta\nu=(\nu_{15}-\nu_{09})=(-1.0\pm0.5)\times10^{-7}$
Hz.  To evaluate the effect on this value of the positional
uncertainty we maximized the difference between the frequency shift
evaluated using Eq.~\ref{eq:poserr} for the 2009 and the 2015
outburst, respectively, obtaining
$\sigma_{\Delta\nu}^{pos}=0.5\times10^{-7}$ Hz. Adding in quadrature
this term to the statistical uncertainty yields an estimate of the
frequency change $\Delta\nu=(-1.0\pm0.7)\times10^{-7}$ Hz, compatible
with zero within less than 2$\sigma$. A spin down rate of
$\mbox{few}\times10^{-15}$ Hz s$^{-1}$ has been observed from all the
four AMSPs for which pulsations were detected during consecutive
outbursts, so far
\citep{hartman2008,hartman2009,patruno2012b,patruno2010b,hartman2011,papitto2011,riggio2011b,patruno2009c}.
It is therefore useful to quote the measured 3$\sigma$ upper limit on
the spin frequency decrease of {\src}
($|\Delta\nu|<3\times10^{-7}$~Hz), and the corresponding limit on the
average rate of spin frequency variation
($|\dot{\nu}|<1.7\times10^{-15}$~Hz~s$^{-1}$).

A more conservative upper limit of the magnitude of the spin
  frequency derivative of {\src} during quiescence was obtained taking
  into account that the pulsar spun up during the $\approx$ 18 d
  elapsed since the 2009 {\xmm} observation and the end of that
  outburst \citep{riggio2011}, and assuming that it did the same
  during the nearly three days occurred since the 2015 outburst
  beginning and the {\xmm} observation. Considering the spin-up rate
derived by \citet{riggio2011}
($\dot\nu_{09}=(1.45\pm0.16)\times10^{-13}$~Hz~s$^{-1}$), we estimated
a frequency increase of
$\Delta\nu^{accr}\simeq(+2.6\pm0.3)\times10^{-7}$ Hz taking place at
the end of the 2009 outburst and at the beginning of the 2015
outburst. Adding this value to the upper limit on the measured
  frequency change, we obtained an upper limit on the spin-down
  rate of the pulsar during quiescence of
  $|\Delta\nu|<|\Delta\nu^{accr}-\Delta\nu|< 6\times10^{-7}$~Hz. The
  corresponding limit on the spin down rate is
  $|\dot{\nu}|<3.5\times10^{-15}$~Hz~s$^{-1}$.

 Assuming that the main torque operating on AMSPs during quiescence
 was due to electromagnetic spin-down, the upper limit on the spin
 down rate can be translated into a constraint on the strength of the
 magnetic dipole moment. Considering the relation derived by
 \citet{spitkovsky2006} for the spin-down of a force-free, rotating
 magnetosphere, one can estimate the NS magnetic dipole moment, $\mu$,
 needed to produce a certain spin down as
\begin{equation}
\mu=\left[\frac{I_* c^3\dot{\nu}}{(1+\sin^2\alpha)(2\pi)^2\nu^3} \right]^{1/2}.
\end{equation}
Here, $\alpha$ is the magnetic inclination angle. Setting
$\alpha=30^{\circ}$, and considering the conservative upper limit
  on the spin down rate estimated earlier
  ($|\dot{\nu}|<3.5\times10^{-15}$~Hz~s$^{-1}$) yielded an upper limit
  of $\mu<3.6\times10^{26}$~G~cm$^3$.

In principle, the torque acting on a pulsar that is propelling away
the inflowing matter through centrifugal inhibition of accretion (the
so-called propeller effect, \citealt{illarionov1975}) is an
alternative explanation of the spin down of AMSP during
quiescence. Assuming that an accretion disk surrounds the pulsar even
during quiescence and is truncated at a radius $R_{in}$ exceeding the
co-rotation radius $R_{c}$, the propeller spin down torque is
$N_{sd}=\dot{M}_d\sqrt{GMR_{in}}[(R_{in}/R_c)^{2/3}-1]$ (see, e.g.,
Eq.~17 of \citealt{papitto2015}, evaluated for completely anelastic
propeller interaction). Assuming $R_{in}=80$~km, a disk mass accretion
rate of $\approx1.5\times10^{-11}$~M$_{\odot}$~yr$^{-1}$ is required
to explain a spin-down rate of $\simeq 3\times10^{-15}$ Hz s$^{-1}$ in
terms of the propeller effect. The propeller luminosity associated to
such a value is $L_{prop}\simeq2.5\times10^{34}$~erg~s$^{-1}$ (see
Eq.~18 of \citealt{papitto2015}). No estimates of the high energy
emission during quiescence are available for {\src}. However, such a
value exceeds by up to two orders of magnitude the X-ray emission
observed from other AMSPs in quiescence, and we deem it as unlikely. A
value of the X-ray luminosity in excess of $\approx 10^{33}${\lumcgs}
would  then be taken as an indication of the presence of
strong outflows from the system, as it was done by
\citet{papitto2014,papitto2015} to interpret the emission of two
transitional millisecond pulsars in the accretion disk state.

We detected three consecutive type-I X-ray bursts during the {\xmm}
observation performed in 2015. Two more were detected by {\igr} but
the much larger time elapsed between them ($\approx$ 3.5 d) indicates
that they were not consecutive. No evidence of burst radius expansion
was detected in the three bursts. The peak flux observed during the
bursts presented here ($\sim1.5\times10^{-8}$~{\fluxcgs} in the
0.5--10 keV band, $\sim3\times10^{-8}$~\fluxcgs in the 3--20 keV) is
not larger than the value reported by \citet{altamirano2010} for the
burst seen during the 2009 outburst ($6.7\times10^{-8}$~\fluxcgs). As
a consequence, the non observation of photospheric radius expansion
during the bursts analyzed here could not yield a tighter constraint
on the source distance than that reported by
\citet[][6.9~kpc]{altamirano2010}. During the 2009 event, eighteen
bursts were detected, each characterized by an exponential decay
timescale of $\simeq10$ s
\citep{altamirano2010,bozzo2010,papitto2010,falanga2011}. The
dependence of the burst recurrence time $t_{rec}$ on the persistent
X-ray flux was studied by \citet{falanga2011}, who found a dependence
$t_{rec}\propto F_{pers}^{-1.1}$. Using the relation plotted in their
Fig.~10, at a bolometric flux of $\simeq1.4\times10^{-9}${\fluxcgs}
(such as that deduced from the simultaneous {\xmm}-{\igr} spectral
modelling, see Table~2), a recurrence time of $\simeq9$ hr is
expected. This is only slightly larger than the recurrence time
observed by {\xmm} in 2015 ($\simeq 8$ hr).

\section{Conclusions}

We have presented an analysis of {\xmm}, {\inte} and {\swift}
observations performed during the outburst detected from {\src} during
early 2015, the second observed from the pulsar after the discovery
outburst in 2009. The outburst profile, spectral and burst properties
were remarkably similar to those observed during the last accretion
event detected in 2009, suggesting that the properties of the
accretion flow did not change much in the two episodes. The frequency
of the coherent signal detected by {\xmm} three days after the
beginning of the outburst was compatible with the value measured by
the same observatory during the 2009 outburst. Therefore, a firm
assessment of the spin evolution of the pulsar during the time elapsed
between the two outbursts was not possible. However, taking into
account the accretion driven spin up observed during the 2009
outburst, and assumed for the 2015 outburst, we derived an upper limit
of $3.5\times10^{-15}$~Hz~s$^{-1}$ on the spin down rate during
quiescence. Electromagnetic spin-down of a NS with a magnetic field
weaker than $3.5\times10^{8}$ G (at the equator of the NS and assuming
a magnetic inclination of $30^{\circ}$) can account for this inferred
spin-down. A magnetic field of the same order has been inferred also
for other AMSPs. Observations of future outbursts will allow to derive
more stringent constraints on the long-term spin evolution of the
pulsar, as well as enabling a first estimate of the orbital period
derivative.

\begin{acknowledgements}

We thank N. Schartel, who accepted the {\xmm} ToO observation in the
Director Discretionary Time, and the XMM–Newton team who performed and
supported this observation. This work is also based on observations
with INTEGRAL, an ESA project with instruments and science data centre
funded by ESA member states (especially the PI countries: Denmark,
France, Germany, Italy, Switzerland, Spain), and with the
participation of Russia and the USA. This work was partly done in the
framework of the grant SGR2014-1073 and AYA2015-71042-P. AP, EB, DFT
and CF acknowledge the International Space Science Institute
(ISSI-Bern) which funded and hosted the international team ``The
disk-magnetosphere interaction around transitional millisecond
pulsars''. AP acknowledges support via an EU Marie Sk\l{}odowska-Curie
fellowship under grant no.~660657-TMSP-H2020-MSCA-IF-2014, and partial
support from “NewCompStar”, COST Action MP1304.  PR acknowledges
contract ASI-INAF I/004/11/0. JJEK acknowledges support from the ESA
research fellowship programme.

\end{acknowledgements}

\bibliographystyle{aa}
\bibliography{biblio}

\end{document}